# Tuning the two-dimensional electron liquid at oxide interfaces by buffer-layer-engineered redox reactions

Yunzhong Chen,[*1] Robert J. Green, [†2,3] Ronny Sutarto,[4] Feizhou He,[4] Søren Linderoth,[1] George A. Sawatzky,[2] and Nini Pryds[1]

[1]*Department of Energy Conversion and Storage, Technical University of Denmark, Risø campus, 4000 Roskilde, Denmark*

[2]*Stewart Blusson Quantum Matter Institute, Department of Physics and Astronomy, University of British Columbia, Vancouver, British Columbia V6T 1Z4, Canada*

[3]*Max Planck Institute for Chemical Physics of Solids, Nothnitzerstraße 40, 01187 Dresden, Germany*

[4]*Canadian Light Source, Saskatoon, Saskatchewan S7N 2V3, Canada*

**ABSTRACT: Polar discontinuities and redox reactions provide alternative paths to create two-dimensional electron liquids (2DELs) at oxide interfaces. Herein, we report high mobility 2DELs at interfaces involving $SrTiO_3$ (STO) achieved using polar $La_{7/8}Sr_{1/8}MnO_3$ (LSMO) buffer layers to manipulate both polarities and redox reactions from disordered**

---

[*]Corresponding Authors. Email:yunc@dtu.dk;

[†]Corresponding Author. Email: rgreen@physics.ubc.ca;

**overlayers grown at room temperature. Using resonant x-ray reflectometry experiments, we quantify redox reactions from oxide overlayers on STO as well as polarity induced electronic reconstruction at epitaxial LSMO/STO interfaces. The analysis reveals how these effects can be combined in a STO/LSMO/disordered film trilayer system to yield high mobility modulation doped 2DELs, where the buffer layer undergoes a partial transformation from perovskite to brownmillerite structure. This uncovered interplay between polar discontinuities and redox reactions via buffer layers provides a new approach for the design of functional oxide interfaces.**

**KEYWORDS:** Complex oxide interfaces, charge transfer, two-dimensional electron liquids, modulation doping.

The interfaces of transition metal oxide heterostructures exhibit a multitude of remarkable properties substantially different from those of their bulk counterparts and conventional semiconductor interfaces. Of particular interest are the highly studied two-dimensional electron liquids (2DELs) at oxide interfaces, such as the perovskite $LaAlO_3$/$SrTiO_3$ (LAO/STO)[1] and the spinel-perovskite, $\gamma$-Al2O3/SrTiO3 (GAO/STO)[2] interfaces. These oxide interfaces with 2DELs occupying $3d$ orbitals show exotic properties, such as magnetism[3], superconductivity[4-6], tunable metal-insulator transitions[7,8], and quantum Hall effect.[9]

The most common approach for the generation of 2DELs at STO-based interfaces is through the epitaxial growth of polar films on nonpolar STO substrates. The subsequent polar discontinuity can induce an electronic reconstruction, forming a 2DEL at the STO side of the interface[1,10]. The paradigmatic example of this approach is the epitaxial LAO/STO heterostructure, where for LAO thicknesses above 4 unit cells (uc) an electronic reconstruction occurs and forms a 2DEL.

However, an important issue for these STO based 2DELs is the presence or absence of oxygen vacancies. For the high growth temperatures (700-800 ˚C) and low oxygen pressures (<$10^{-4}$ mbar) often used for the growth of epitaxial LAO films on STO, oxygen vacancies can be formed in the STO, generating free carriers in addition to those due to the electronic reconstruction. Extensive investigations into oxygen vacancies have been performed for the epitaxial LAO/STO interface[11-19], and indeed care must be taken to eliminate oxygen vacancies and realize a pure electronic reconstruction. Typically this is achieved by post-annealing the sample under high oxygen pressures[11,12,16,20]. However, such an approach generally leads to a reduced electron mobility of the 2DEL (on the order of ~1000 cm$^2$V$^{-1}$s$^{-1}$ at 2 K) compared to non-annealed samples or their bulk counterpart (~20000 cm$^2$V$^{-1}$s$^{-1}$ at 2 K).

Alternatively, 2DELs have also been generated by the room temperature growth of oxide films on STO, where redox reactions lead to oxygen vacancies in the STO directly at the interface, and subsequent free carriers. Notably, strong redox reactions can occur when growing perovskite oxides (*AB*O$_3$) capping films having B site cations of Al, Ti, Zr, and Hf or other elements with a heat of metal oxide formation lower than -350 kJ per mole of oxygen, and a work function, $\varphi$, in the range of 3.75 eV $< \varphi <$ 4.4 eV[21,22]. The room temperature synthesis generally leads to disordered (d) or amorphous-like film growth, and 2DELs have been observed for heterostructures of STO capped with disordered LaAlO3 (*d*-LAO), disordered yttria-stabilized zirconia (*d*-YSZ), and disordered strontium titanate (*d*-STO)[16,21,23-25]. This approach is applicable to not only (001)-oriented but also (110) and (111)-oriented STO crystals[25]. Nevertheless, these oxygen vacancy dominated 2DELs are characterized by a pronounced carrier freeze out effect, and it remains also challenging to control their formation without a compromise in the properties[16,21,23-25].

Recently it was found that, without any post-annealing, the oxygen vacancies on the STO side of $d$-LAO/STO interfaces were suppressed strongly by inserting a single-unit-cell (uc) polar La$_{7/8}$Sr$_{1/8}$MnO$_3$ (LSMO) buffer at the interface[22]. In this case, the LSMO buffer acts as an electron sink during the formation of interfacial 2DELs, since the partially-filled Mn $3d$ $e_g$ subbands is often lower than the Fermi level of Ti $3d$ $t_{2g}$ bands. This has led to a charge-transfer-induced modulation doping and boosts the 2DEL electron mobility more than 20 times of $d$-LAO/STO[23]. However, a new puzzle arises, as the total amount of transferred electrons at the d-LAO/LSMO(1uc)/STO interface is found to be approximately 1 e/uc, which is far above the 0.5 e/uc as normally expected by a pure electronic reconstruction scenario considering the polar nature of the buffer layer. More information is clearly needed on the detailed atomic and electronic structures of such promising buffered interfaces to fully understand the origin of the high mobility 2DEL.

In this Letter, we studied LSMO-buffered interfaces between STO substrates and three different disordered oxide films, $d$-LAO, $d$-YSZ, and $d$-STO, in order to gain more insight into the buffer layer induced high mobility. We find that the LSMO buffer layer considerably suppresses the formation of oxygen vacancies on the STO side of the interface in all three types of heterostructures at a suitable thickness. Additionally, we perform an in-depth analysis of the atomic and electronic structures of the buried interfaces using resonant x-ray reflectometry. We find that a combination of a polarity induced electronic reconstruction of the epitaxial LSMO and redox reactions at the interface of the disordered overlayer and LSMO, which yields the high mobility, modulation doped 2DEL. The redox reaction is strongly confined at the interface between the LSMO buffer and capping films, which can also explain the large amount of electrons previously found in the buffer layers for $d$-LAO/LSMO/STO[23]. The uncovered interplay between polar discontinuities and redox reactions provides a new landscape for the design of functional oxide interfaces.

The heterostructures, as illustrated in Fig. 1(a), were grown by pulsed laser deposition (PLD) where the thickness of LSMO buffer layer, $t$, is well controlled with precision on the scale of uc. Figures 1(b)-(d) show the typical temperature dependent sheet resistance, $R$s, of $d$-LAO/LSMO/STO, $d$-YSZ/LSMO/STO, and $d$-STO/LSMO/STO heterostructures for various thicknesses $t$ of the LSMO buffer layer. In the absence of a buffer layer, all three heterostructures show a metallic behavior with comparable electron density, $n_s$, in the order of $1.0 \times 10^{14}$ cm$^{-2}$ at 295 K, where the conduction results primarily from the interfacial redox reaction, i.e. formation of oxygen vacancies on the STO side[16,21]. The conduction exhibits typically a carrier freezing-out effect in the temperature dependence of the sheet carrier density at $T$<100 K (supporting information) due to the presence of impurity states in the band gap just below the bottom of STO conduction band[16,26]. The temperature-dependent carrier density can be expressed as $n_s(T) = n_s(5$ K$) + n_0 \exp(-E_a/k_B T)$. Here $n_0$ is the saturated carrier density at high temperature ($T \geq 100$ K), $E_a$ is the thermal activation energy (the energy levels of the donor states below the conduction band minimum) and $k_B$ is the Boltzmann constant. The non-buffered samples exhibit an $E_a$ of 7-8 meV.

Upon the insertion of LSMO, the interface conduction in all the three heterostructures is strongly suppressed. However, the buffer layer is found to have two very different effects, depending on the thickness and the identity of the disordered overlayer: First, for all $d$-STO/LSMO/STO samples and those of $d$-YSZ/LSMO/STO with $t \leq 4$ uc, the ns decreases gradually upon increasing $t$ (Fig. 2(a)), and the carrier freezing-out phenomenon is always present (see supplementary information). The carrier freezing-out phenomenon results from defect states induced by oxygen vacancies on the STO side.[26] Its presence, therefore, indicates that the conduction in these samples is still dominated by the interface redox reaction although with fewer STO oxygen vacancies formed for thicker LSMO buffer layers because of their lower $n_s$ and smaller $E_a$. Interestingly, we observed an approximately linear dependence of $E_a$ with respect to $t$, where the slope is about -0.7 meV/uc (Fig. 2(c)). This suggests that the redox-induced conduction, if not interrupted by other electronic effects, can persist up to $t$=7meV/(0.7 meV/uc)=10 uc, as observed in the $d$-STO/LSMO/STO case. However, a second type of remarkable buffer layer effect is observed in $d$-LAO/LSMO/STO and $d$-YSZ/LSMO/STO at $t$ = 1 uc and 4 uc$\leq t \leq$6 uc, respectively, where the carrier freezing-out effect is suppressed completely and also a large enhancement in the carrier mobility is found (Fig. 2b). Upon increasing $t$ further for these samples, a sharp metal-insulator transition is observed, which suggests that the thick LSMO buffer layer blocks completely the formation of interface 2DEL, i.e. no electrons transferred to the Ti 3$d$ $t_{2g}$.

The high mobility conduction observed in $d$-YSZ/LSMO/STO is similar to that reported earlier for $d$-LAO/LSMO/STO[23], but its presence at a much thicker buffer layer thickness suggests that a different mechanism could be at play. To gain more insight into the detailed atomic and electronic structures of the interface regions in the $d$-YSZ/LSMO/STO system, we performed resonant x-ray reflectometry (RXR) experiments on a series of samples. RXR is a non-destructive technique which can probe elemental and valence concentration depth profiles, ideally suited to the study of oxide heterostructures[27-29]. The RXR experiments were performed using the in-vacuum 4-circle diffractometer at the Resonant Elastic and Inelastic X-ray Scattering (REIXS) 10ID-2 beamline of the Canadian Light Source (CLS)[30].

In Fig. 3(a-c), we show the results of RXR analysis on a controlled series of samples: $d$-YSZ/STO, epitaxial LSMO/STO, and an optimally buffered $d$-YSZ/LSMO/STO heterostructure, respectively. In the upper portion of the figure, we schematically show the valence profiling results from the RXR, while in the bottom panels we show the quantitative extracted element and valence profiles which include interface roughness. These concentration profiles are offset into anions (top), A-site cations (middle), and B-site cations (lower) for clarity.

For the $d$-YSZ/STO sample in Fig. 3(a), our main finding is the presence of a high concentration of $Ti^{3+}$ at the interface ($\sim$2.3 $e$ per square unit cell, spread primarily over two unit cells near the interface). RXR is very sensitive to the valence of the Ti at the interface through the resonance of the atomic form factor. This is exhibited in Fig. 3(d), where we show an experimental fixed angle reflectivity scan taken at $\Theta = 30°$ across the resonance, along with our model fit result which includes the interfacial $Ti^{3+}$. Shown also is a model calculation with no $Ti^{3+}$ at the interface, which strongly deviates from the experimental result. Additionally, as shown in the lower panel of Fig. 3(a) we observe a dip in the oxygen density profile at the interface, consistent with the transport measurements which indicated that the $Ti^{3+}$ in this case arises from redox reactions leading to oxygen vacancies ($V_O$) at the interface. We find a total of 1.6±0.6 $V_O$, which is consistent with the amount of $Ti^{3+}$ (see supplemental material for full datasets). Note that the charge quantity found with RXR is larger than that of the transport measurements of Fig. 2(a), indicative of a significant charge portion which is localized, similar to that found for crystalline LAO/STO heterostructures[31].

In Fig. 3(b), we show RXR results for an epitaxial, insulating LSMO film on STO. Here, no $Ti^{3+}$ is observed (consistent with lack of conductivity in transport) and instead the polar discontinuity which is present in this case is accommodated by a layer of $Mn^{2+}$ at the interface (as shown in Fig.

3(c)), amounting to 0.60 $Mn^{2+}$ ions per square unit cell. A small $Mn^{2+}$ portion of 0.40 $Mn^{2+}$ ions per unit cell is also present at the surface of the LSMO, in this case due to slight surface structure degradation, also evidenced by the off stoichiometry in the La concentration profile (such surface $Mn^{2+}$ is commonly found in uncapped LSMO films[32]). This electronic reconstruction in the form of interfacial $Mn^{2+}$ rather than $Ti^{3+}$ is consistent with other recent reports of insulating behavior at $LaMnO_3/SrTiO_3$ polar/non-polar interfaces[33].

Figure 3(c) shows the RXR results for the optimally buffered d-YSZ/LSMO/STO sample, which combines the redox and polarity effects of the other two systems. Here, the $Ti^{3+}$ is below detection limits, consistent with the reduced carrier densities found in the transport measurements (on the order of 0.01 e/uc), indicating that the redox reactions are strongly suppressed in the STO substrate. However, through the Mn $L_{2,3}$ resonance spectra (e.g. lower curves in Fig. 3(e), full dataset in Supplemental Material), effects of both redox reactions and electronic reconstruction are detected in the buffer layer. First, as shown in Fig. 3(c), approximately 0.39 $Mn^{2+}$ ions per square unit cell are present at the STO/LSMO interface, similar to the case of the LSMO/STO sample in Fig. 3(b). This charge again likely arises due to the polarity induced electronic reconstruction as in pure LSMO/STO without a capping film. However, there is substantial $Mn^{2+}$ also found at the upper d-YSZ/LSMO interface, considerably more than the uncapped LSMO/STO sample of Fig. 3(b). Here we identify both octahedrally $O_h$ and tetrahedrally $T_d$ coordinated $Mn^{2+}$, along with a dip in the oxygen concentration [supporting information], suggesting the formation of the oxygen deficient brownmillerite structure in a thin layer at the interface. Such a structural transformation has been observed before for LAO and STO film growth on LSMO buffer layers[34], though at a high growth temperature above 400˚C. Uniquely, our room temperature growth limits the brownmillerite to the upper 2uc of the LSMO. This results in an unprecedented charge reconstruction in the heterostructure: the LSMO buffer layer exhibits an electronic reconstruction at the LSMO/STO interface, followed by a section of pure LSMO, and then finally with a brownmillerite-like section at the interface to the *d*-YSZ.

The detailed electronic and atomic structures uncovered for the LSMO buffer layer in *d*-YSZ/LSMO/STO offers further insight into the unique transport properties reported in Figs. 1 and 2. As is evident for the uncapped LSMO/STO, the electronic reconstruction does not create a conducting interface, because the electron affinity of the $Mn^{3.125+}$ of the LSMO is stronger than that of the $Ti^{4+}$ of the STO, leading the reconstructed charge to be localized in the form of $Mn^{2+}$. However, the presence of the redox-induced brownmillerite layer at the upper *d*-YSZ/LSMO interface seems crucial for yielding further charge transfer from the $Mn^{2+}$ into the Ti of the STO – when the LSMO is thicker than 6uc where the brownmillerite is further away from the LSMO/STO interface, the system becomes insulating. The charge transfer between $Mn^{2+}$ and STO is determined by the band alignment between Mn $3d$ $e_g$ electrons and the Ti $3d$ $t_{2g}$ electrons, where the Mn $3d$ $e_g$ levels depend strongly on the strain-induced lattice distortion and the potential distortion due to the brownmillerite layer at the *d*-YSZ/LSMO interface (similar to LAO surface oxygen vacancies for the LAO/STO system[35]). It has been previously revealed that when the LSMO is strongly strained on STO (such as at $t \leq$ 6uc), orbital reconstructions result in a situation that the Mn $e_g$ levels become comparable to or even higher than the Ti $t_{2g}$ electron level[36], in this case, charge transfer could occur from $Mn^{2+}$ to STO (as indicated by the arrow in the upper schematic of Fig. 3(c)). When LSMO becomes thick (such as $t>$ 6uc), the Mn $e_g$ levels turn lower than the Ti $t_{2g}$ level, therefore, there will be no charge transfer from $Mn^{2+}$ to STO, i.e. no formation of 2DEL. Such an interpretation agrees also with the optimal buffer layer thickness range of 4-6 uc as found in the transport measurements. Below 4 uc, the polar potential of the LSMO is below the threshold for electronic reconstruction, and the conduction in STO remains dominated by the redox reaction. Above 6 uc, besides lattice distortions[36], the *d*-YSZ/LSMO interface which hosts the brownmillerite layer, is further from the LSMO/STO interface, and the potential is no longer a large enough effect to alter the Ti and Mn electron affinities at the lower interface. The combination of electronic reconstruction at the bottom interface and redox reaction at the other in $d$-YSZ/LSMO/STO heterostructures may also occur in the $d$-LAO/LSMO/STO system[23]. However, since the buffer layer is only one unit cell, it remains challenging to distinguish these two effects. As for the $d$-STO/LSMO/STO system, it differs from the other two systems in two aspects. Firstly, the conduction band of $d$-STO is on the same level or lower than that of crystalline STO; Secondly, the redox reactions persist even when LSMO is quite thick (10 uc), where the Mn $e_g$ band will be always lower than that of the Ti $t_{2g}$ band. Consequently, a pure electronic charge transfer from $Mn^{2+}$ to STO thus the high mobility 2DEL is not achievable in $d$-STO/LSMO/STO.

In summary, our study reveals a complex interplay between polarity and redox reactions in determining the properties of 2DEL at STO-based interface. Similar to LSMO buffered $d$-LAO/STO reported earlier, high mobility enhancement is also achieved for LSMO buffered $d$-YSZ/STO interfaces. In this case, redox reactions are confined to the LSMO buffer and lead to a partial transformation of the manganite to a brownmillerite structure. While pure LSMO reconstructs to form an insulating, $Mn^{2+}$ containing interface, the brownmillerite structural transformation with disordered overlayer leads to charge transfer into the STO, forming a high mobility, modulation-doped 2DEL. This buffer layer approach which engineers both polar discontinuities and redox reactions provides a promising new landscape for the design of functional oxide interfaces.

**METHODS**

**Sample growth and Characterizations.** The heterostructures, as illustrated in Fig. 1(a), were grown by pulsed laser deposition (uence of 4 J/cm$^2$, 1Hz) on (001)-oriented STO substrates with single TiO$_2$ layer termination. During film growth, the epitaxial LSMO layer, which was accurately controlled on a unit cell scale by monitoring in situ reflection high-energy electron diffraction (RHEED) intensity oscillations under optimized condition, was first deposited at 600 ˚C and $1\times10^{-4}$ mbar of O$_2$, then slowly cooled to room temperature at the deposition pressure before switching to the in situ deposition of $d$-LAO, $d$-STO, and $d$-YSZ films. The final film deposition near room temperature (below 50 ˚C) results in the formation of disordered or amorphous-like capping films (the film thickness is in the range of 4-9 nm). Electrical characterization was performed using a 4-probe Van der Pauw method with ultrasonically wire-bonded aluminum wires as electrodes.

**Resonant x-ray reflectometry (RXR) measurements.** The experiments were carried out at the REIXS beamline of the Canadian Light Source, at 300 K in an ultra-high vacuum environment ($< 10^{-9}$ mbar). The modelling of the RXR was completed with the program QUAD. A multi-variable Levenberg-Marquardt algorithm was used to fit the simulation to the experimental data [supporting information]. Extended RXR datasets and corresponding fits which are either energy scans through particular resonances at fixed reflection angles, or reflection angle scans at fixed energies are also detailed in the supporting information.

**Acknowledgement.** We thank the technical assistance from J. Geyti, the financial support from the Denmark Innovation Fund, from NSERC and the Max Planck-UBC Centre for Quantum Materials, Canada Foundation for Innovation, the National Research Council of Canada, the Canadian Institutes of Health Research, the Government of Saskatchewan, Western Economic Diversification Canada, and the University of Saskatchewan.

Y.Z.C. and R. J. G. contributed equally to this work

**Supporting Information Available**. This material is available free of charge via the Internet at <u>http://pubs.acs.org</u>.

**FIGURE CAPTIONS**:

**Fig. 1** (a) Sketch of the manganite-buffered disordered/crystalline heterostructures. (b)-(d) The temperature dependent sheet resistance for $d$-LAO/LSMO/STO, $d$-YSZ/LSMO/STO, and $d$-STO/LSMO/STO respectively, for varying thicknesses $t$ of the LSMO buffer layer.

**Fig.2.** Effect of buffer layer thickness, $t$, on the transport properties. (a) Carrier densities at 295 K (b) Mobilities at 2.5 K. (c) Extracted activation energies.

**Fig.3.** Resonant x-ray reflectometry results. (a-c, upper) Schematic depictions of the Ti and Mn valence changes at the interfaces for (a) $d$-YSZ/STO, (b) LSMO/STO, and (c) $d$-YSZ/LSMO/STO heterostructures. (a-c, lower) Element and valence concentration profiles for the three samples, offset into anions, A-site cations, and B-site cations for clarity. The interface valence changes are labeled for the B-site densities. (d-e) RXR experimental data and fits exhibiting the sensitivity to the interface valence changes for (d) $d$-YSZ/STO and (e) LSMO/STO and $d$-YSZ/LSMO/STO. In each case, a model spectrum with no interface valence change is shown for comparison to the best fit containing the valence change. Full datasets and fits can be found in the supplemental material.

**Table of Contents Graphic**

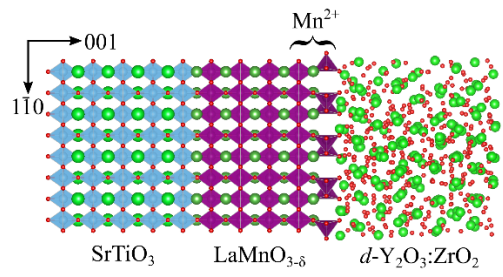

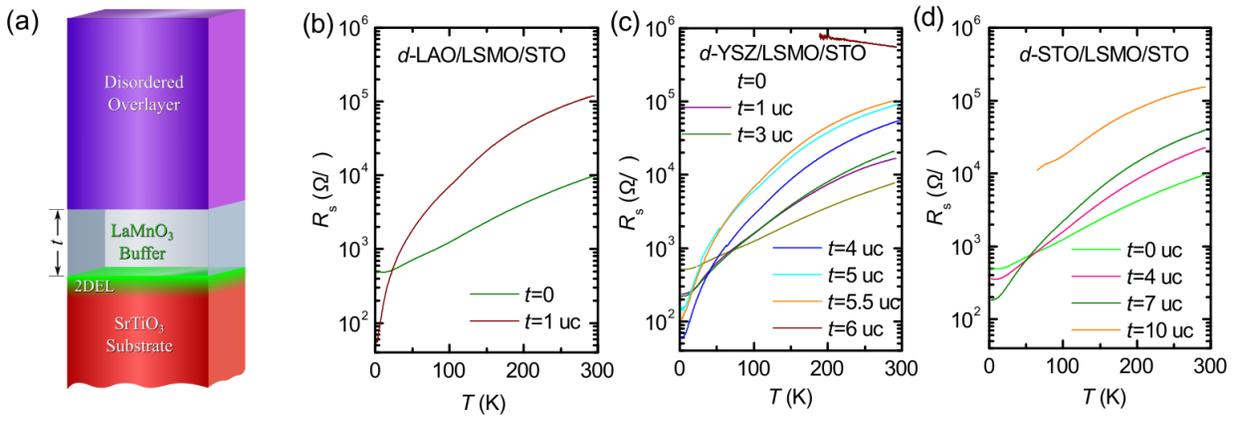

**Fig.1**

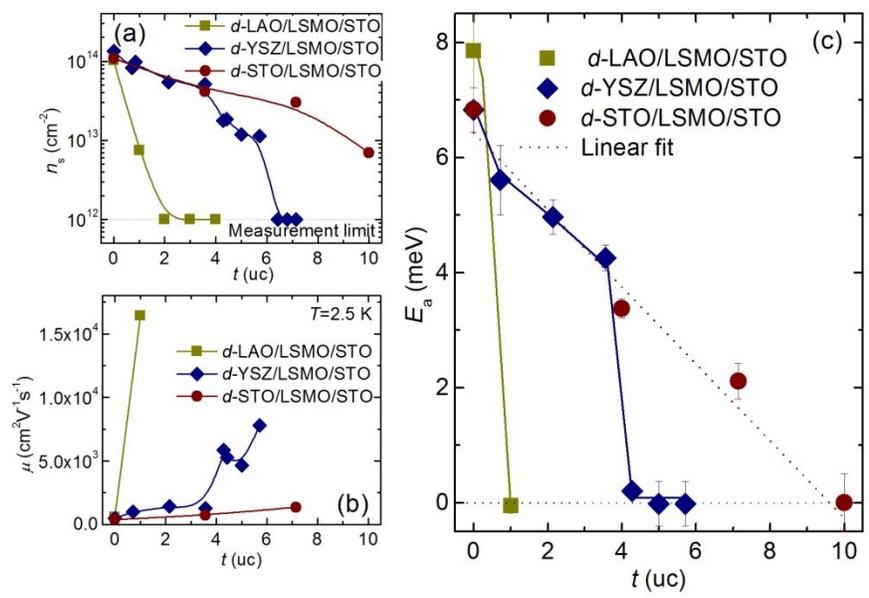

**Fig.2**

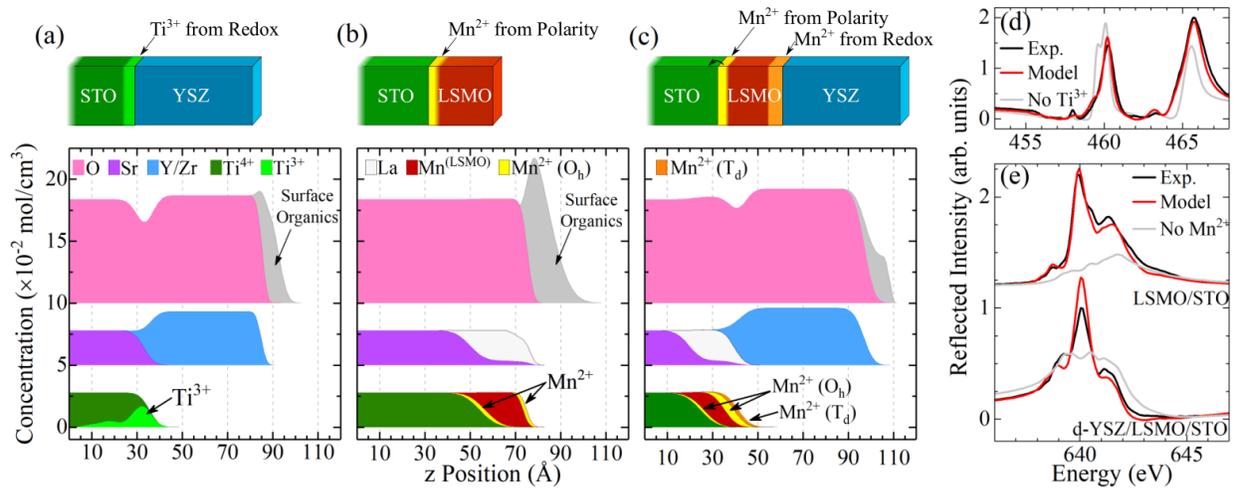

**Fig.3**